\RequirePackage{lineno}
\documentclass[twocolumn,superscriptaddress,preprintnumbers,amsmath,amssymb,prc]{revtex4}  
\usepackage{graphicx}
\usepackage{dcolumn}
\usepackage{bm}
\usepackage{float}
\usepackage{color}
\usepackage{CJK}
\usepackage[colorlinks,linkcolor=blue,urlcolor=blue,citecolor=blue]{hyperref}
\usepackage{isotope}
\usepackage[normalem]{ulem}
\renewcommand{\sout}{\bgroup \color{red} \ULdepth=-0.5ex \ULset}

\usepackage{amsmath}
\usepackage{cases}
\usepackage{amssymb}
\usepackage{eqnalign}

\begin{document}

\title{Search for jet quenching effects on the plain jet mass in Pb+Pb collisions at the LHC with a multiphase transport model}

\author{Xiang-Pan Duan}
\affiliation{Key Laboratory of Nuclear Physics and Ion-beam Application~(MOE), Institute of Modern Physics, Fudan University, Shanghai $200433$, China}
\affiliation{Shanghai Research Center for Theoretical Nuclear Physics, NSFC and Fudan University, Shanghai $200438$, China}

\author{Guo-Liang Ma}
\email{glma@fudan.edu.cn}
\affiliation{Key Laboratory of Nuclear Physics and Ion-beam Application~(MOE), Institute of Modern Physics, Fudan University, Shanghai $200433$, China}
\affiliation{Shanghai Research Center for Theoretical Nuclear Physics, NSFC and Fudan University, Shanghai $200438$, China}

\begin{abstract}
{
The plain jet mass distributions of reconstructed jets are investigated in p+p and 0-10$\%$ most central Pb+Pb collisions at $\sqrt{s_{\rm NN}} = 2.76~{\rm TeV}$ using a dynamical multiphase transport model with a string melting mechanism. It is observed that the mean charged jet mass increases with increasing jet transverse momentum and jet radius in central Pb+Pb collisions. It is demonstrated that the plain jet mass of reconstructed partonic jet is shifted to a higher value after the evolution of partonic stage due to jet quenching in central Pb+Pb collisions. However, the jet mass shift effect is strongly weakened by non-perturbative effects from hadronization and hadron rescatterings. This makes it difficult to observe significant hot medium modification effects on the plain jet mass distribution in the final state of relativistic heavy-ion collisions.
}
\end{abstract}

\maketitle

\section{Introduction}\label{1}

Jet is produced by hard quantum chromodynamics (QCD) scatterings as a high energy parton shower with high virtuality~\cite{Sterman:1977wj,Feynman:1978dt, Renk_PRC79(2009)}, but finally evolves into a hadron shower due to the QCD confinement dominated by non-perturbative QCD physics~\cite{Dasgupta_JHEP055(2008)}. When jet passes a hot and dense QCD medium, it will lose energy due to jet-medium interactions~\cite{QGY_PRL103(2009),Shi:2018izg,Zhang:2017ebx}. This phenomenon is named as jet quenching~\cite{Gyulassy_PLB243(1990),WXN_PRL68(1992),CMS_PRC84(2011),QGY_IJMPE24(2015),CSS_RPP84(2021),Xu:2014tda}, which makes jet as an important hard probe to detect the properties of quark-gluon plasma (QGP) at the Relativistic Heavy Ion Collider (RHIC)~\cite{BRAHMS_PRC91(2003),STAR_PRL91(2003)172302,PHOBOS_PLB578(2004),STAR_NPA757(2005),PHENIX_NPA757(2005),Shen:2020mgh,Wu:2021xgu} and the Large Hadron Collider (LHC)~\cite{ATLAS_PRL105(2010),ALICE_PLB696(2011),CMS_EPJC72(2012),Song:2017wtw}. Both the modification of the peaks in two-particle azimuthal correlation~\cite{STAR_PRL90(2003),STAR_PRL91(2003)172302,Wong:2008zzb} and the suppression of nuclear modification factor $R_{\rm{AA}}$ of high transverse momentum particles~\cite{STAR_PRL91(2003)172302,STAR_PRL91(2003)072304,PHENIX_PRL91(2003),Betz:2015oia} are explained as the effects of jet quenching. Because jet is a collimated spray of particles with high transverse momenta, it can be fully reconstructed based on the final state hadrons in the experiment. Recently, many reconstructed jet observables, e.g. dijet transverse momentum asymmetry $A_{\rm{J}}$~\cite{ATLAS_PRL105(2010),QGY_PRL106(2011),MGL_PRC87(2013),Chen_PLB782(2018)}, jet charge~\cite{ATLAS:2015rlw,Krohn:2012fg,Chen:2019gqo}, jet shape~\cite{Chang:2016gjp,KunnawalkamElayavalli:2017hxo,Wan:2018zpq}, and jet structures~\cite{CMS_JHEP87(2012),MGL_PRC88(2013),MGL_PRC89(2014),CNB_PLB801(2020),Luo:2021hoo,Yan:2020zrz}, have been studied to explore the very hot QGP matter created in the relativistic heavy-ion collisions at the LHC.

The jet mass, as another observed quantity of jet properties, has been widely studied on both experimental and theoretical aspects~\cite{ATLAS_JHEP128(2012),CMS_JHEP90(2013),ALICE_PLB776(2018),STAR:2021lvw,Majumder_PRC93(2016),Kunnawalkam_JHEP141(2017),KZB_JHEP10(2018),Idilbi_JKPS73(2018),Park_NPA982(2019),Balsiger_JHEP04(2019),Casalderrey_JHEP44(2020)}.
The jet mass can be calculated as follows~\cite{ALICE_PLB776(2018)},
\begin{equation}\label{eq.1}
M_{\rm {jet}} = \sqrt{{E}^{2}-{p}_{\rm {T}}^{2}-{p}_{\rm {z}}^{2}},
\end{equation}
where $E$, $p_{\rm {T}}$ and $p_{\rm {z}}$ are the jet energy, jet transverse momentum and jet longitudinal momentum, which depend on jet constituents' four momenta for a reconstructed jet. For a quark-initiated or gluon-initiated jet, after emitting many gluons, the ratio of the squared jet mass to the jet energy can be approximately simplified as~\cite{Larkoski:2017fip,Marzani:2019hun},
\begin{equation}\label{eq.2}
M_{\rm {jet}}^{2}/E^{2} = \sum_{i=gluon} z_{i} \theta_{i}^{2},
\end{equation}
where $z_i$ is the energy fraction of the $i^{th}$ gluon and $\theta_{i}$ is the angle of the $i^{th}$ gluon to the initial quark or gluon jet. Thus, the jet mass can be considered as a prototype of the jet shape. The two competing effects in jet quenching, jet energy loss and medium response, have been argued to transform the jet mass distribution differently~\cite{Kunnawalkam_JHEP141(2017),Park_NPA982(2019),Casalderrey_JHEP44(2020)}. According to Eq.~(\ref{eq.2}), the jet energy loss due to jet quenching leads to a decrease of jet mass, however, the medium response helps jet mass to rise to a higher value because more soft particles with large $\theta$ angles are involved in the jet cone due to jet-medium interactions. Therefore, the jet mass with the medium response is expected to be larger than that one without medium response. In the recent ALICE measurement~\cite{ALICE_PLB776(2018)}, the plain charged jet mass distribution in central Pb+Pb collisions has been observed to be in agreement with that in p+Pb collisions within experimental uncertainties, where the experimental data of jet mass are also compared with the quenching models and non-quenching models. The result from non-quenching models is consistent with the experimental data in p+Pb collisions, however, the quenching models with medium response overestimate the mean value of plain jet mass distribution in Pb+Pb collisions~\cite{ALICE_PLB776(2018)}. 

On the other hand, the jet mass is predicted to be sensitive to both perturbative and non-perturbative QCD effect~\cite{ATLAS_JHEP128(2012),CMS_JHEP90(2013),STAR:2021lvw,Majumder_PRC93(2016),Idilbi_JKPS73(2018),Balsiger_JHEP04(2019)}. For example, the hadronization of partons~\cite{Field:1977fa,Tan:2005rxa,Yu:2001zq}, as a non-perturbative effect, has been experimentally found to increase the jet mass~\cite{STAR:2021lvw}. Because the virtuality of the jet is calculated to associate with the jet mass~\cite{Majumder_PRC93(2016)}, the virtuality will be decreased with interacting with the QGP matter. This means that parton interactions cause the jet mass depletion along with jet energy loss. However, it is also argued that there could be a temporary virtuality increase in this process, which leads to a temporary increase of jet mass. To distinguish between perturbative and non-perturbative QCD effects, an effective way is to use jet grooming techniques to focus on jet substructure~\cite{Larkoski:2017fip,Marzani:2019hun}. For example, the soft drop grooming can drop both soft and collinear emissions in Lund diagram for jet shower~\cite{Larkoski:2014wba}. Therefore, the groomed mass is considered to be favorable to remove the background from the underlying event and pile up from jet showers. A hint of increased probability to produce jets with large groomed jet mass has been observed by the CMS Collaboration~\cite{CMS:2018fof}, which could originate from medium-induced radiation at a large angle from the jet axis.

In this study, we focus on different dynamical effects on the characteristics of the ungroomed jet mass, also known as plain jet mass, in relativistic heavy-ion collisions, by using a dynamical transport model including parton interactions, hadronization, and hadronic rescatterings. Please note that unless otherwise stated,  the jet mass refers to the plain jet mass in the following. We calculate the jet mass distributions in p+p and 0-10$\%$ most central Pb+Pb collisions at $\sqrt{s_{\rm NN}} = 2.76~{\rm TeV}$ using a multiphase transport (AMPT) model with string melting mechanism~\cite{LZW_PRC72(2005),Lin:2021mdn}, where the effects of different dynamical evolution stages on jet mass are considered.

The paper is organized as follows. In Sec.~\ref{2}, we briefly review the basic framework of the AMPT model with a jet-triggering technique and introduce the method of jet reconstruction and Bayesian unfolding technique. Comparisons of charged jet mass between with and without jet quenching in most central Pb+Pb collisions is presented in Sec.~\ref{3.1}. We further discuss about the effects of different dynamical evolution stages on the jet mass distribution in Sec.~\ref{3.2}. Finally, a summary is given in Section.~\ref{4}.

\section{Methodology}\label{2}

\subsection{The AMPT model with triggered dijet}
The AMPT model with string melting mechanism, which is widely applied in the field of relativistic heavy-ion collisions, includes four main stages: the initial condition generated by the heavy ion jet interaction generator (HIJING) model~\cite{WXN_PRD44(1991),Gyulassy_CPCo83(1994)} where the nuclear shadowing effect is included via an impact-parameter-dependent but $Q^2$(and flavor)-independent parameterization, parton interactions simulated by Zhang’s parton cascade (ZPC) model~\cite{ZB_CPCo109(1998)}, hadronization performed by a simple quark coalescence model and hadronic rescatterings described by a relativistic transport (ART) model~\cite{LBA_PRC52(1995)}. In the string melting mechanism, all parent hadrons from the fragmentation of minijets in hard process and excited strings in soft process are melted into partons according to its constituted quarks, which provides us an initial state of partonic matter. The ZPC model involves two-body elastic scatterings among partons and the parton cross section based on the leading order pQCD gluon-gluon interaction associates with the strong coupling constant and the Debye screening mass. The nearest freeze-out partons are combined into mesons or baryons via a simple quark coalescence model. The three-momentum is conserved during the coalescence. The species of formed hadrons are determined by the flavor and invariant mass of combined partons. The simple quark coalescence model includes the formation of all mesons and baryons listed in the HIJING program. The final state hadronic rescatterings including elastic and inelastic scatterings and resonance decays are simulated by the ART model .

To study the jet quenching effect, a triggered high-$p_{\rm{T}}$ dijet is embedded into the initial condition using the jet-triggering technique of the HIJING model. Several hard QCD processes have been used to produce the initial dijet, which consists of $g+g\rightarrow g+g$, $g+g\rightarrow q+\bar{q}$, $q+g\rightarrow q+g$, $q+\bar{q}\rightarrow g+g$, $q_{1}+q_{2}\rightarrow q_{1}+q_{2}$, and $q_{1}+\bar{q_{1}}\rightarrow q_{2}+\bar{q_{2}}$ ~\cite{Sjostrand_CPCo82(1994)}. Both initial and final radiations of jet showers are considered. In the string melting mechanism, the primordial hadrons that would be produced from the excited Lund strings, minijet and jet partons in the HIJING model are melted into primordial quarks and antiquarks according to the flavor and spin structures of their valence quarks.  Under the above initial conditions, a partonic plasma with a dijet is produced.  A parton cross section, $3~\rm mb$, is performed to effectively simulate the elastic interactions between jet shower partons and medium partons in Pb + Pb collisions at $\sqrt{s_{\rm NN}} = 2.76~{\rm TeV}$. Because our simulation is a full parton cascade, the elastic scatterings between recoiled partons are included automatically. By setting the parton interaction cross section as $0~\rm mb$, we also can turn off both parton cascade and jet-QGP interactions. The jet parton shower is converted to a jet hadron shower through the hadronization of coalescence, when all partons freeze out. Note that if turning off parton cascade, our hadronization of coalescence is actually as same as the hadronization of Lund string fragmentation, since the simple quark coalescence model combines nearest partons into hadrons~\cite{LZW_PRC72(2005)}. In the final state, the jet hadron shower loses its energy in the hadronic phase which is simulated in the stage of hadronic rescatterings. In principle, the parton cross section should be energy-dependent, even with respect to the properties of the surrounding medium. However, our previous studies show that the AMPT model with a proper parton cross section has successfully described many experimental observables, including not only bulk properties~\cite{Lin:2014tya,Ma:2016fve,He:2017tla,Lin:2021mdn} but also reconstructed jet observables, such as $\gamma$-jet imbalance~\cite{MGL_PLB724(2013)}, dijet asymmetry~\cite{MGL_PRC87(2013)}, jet fragmentation function~\cite{MGL_PRC88(2013),Ma:2014PRC896,Duan:2022bew}, jet shape~\cite{MGL_PRC89(2014)}, jet anisotropies~\cite{Nie:2014pla}, jet transport coefficient~\cite{Zhou:2019gqk}, and the redistribution of lost energy~\cite{Gao:2018PRC97,Luo:2022EPJC82}. Therefore, we continue to use the constant parton cross section in this study.

\subsection{Jet reconstruction}
To study the mass of reconstructed jets, our kinematic cuts are taken as same as the ALICE experiment which measured the mass of the charged jets without using grooming algorithms~\cite{ALICE_PLB776(2018)}. These track particles are required to be with the transverse momentum of $p_{\rm T,trk}> 0.15~\rm{GeV}$/$c$ and the pseudorapidity within $\left|\eta_{\text trk}\right| < 0.9$. The jets are clustered with anti-$k_{\rm T}$ algorithm~\cite{Cacciari_JHEP063(2008)}, and the background density is estimated with the $k_{\rm T}$ algorithm in Pb+Pb collisions~\cite{Cacciari_PLB641(2006)} using the FastJet package~\cite{Cacciari_EPJC72(2012)}. The transverse momentum of reconstructed jets is demanded to be larger than $60~\rm{GeV}$/$c$ for charged jets and $100~\rm{GeV}$/$c$ for full jets in our calculations. The E-recombination scheme is performed to reconstruct jets with three jet radii $R$ = 0.2, 0.3, and 0.4, respectively. The jets within a pseudorapidity window of $\left|\eta_{\text jet}\right| < 0.9 - R$ are used to study the jet mass distribution. Our results are specifically compared with experimental data for jet radius $R$ = 0.4. To remove the jet background in  Pb + Pb collisions, an area-based background subtraction method is used to obtain the transverse momentum of background-subtracted jet~\cite{Cacciari_PLB659(2008)},
\begin{equation}\label{eq.3}
p_{{\rm T,\text jet}}^{\rm \text sub} = p_{{\rm T,\text jet}}^{\rm \text raw}-\rho \cdot A_{\rm \text jet}, 
\end{equation}
where $A_{\text jet}$ is the jet area calculated by ghosts with an area of 0.005~\cite{Cacciari_JHEP005(2008)}. The $\rho$ is the background transverse momentum density~\cite{Cacciari_PLB659(2008)}, which is defined as,
\begin{equation}\label{eq.4}
\rho = \operatorname{median}\left[\left\{\frac{p_{\rm {T,j}}}{A_{\rm j}}\right\}\right],
\end{equation}
where $A_{\rm j}$ and $p_{\rm T,j}$ are the area and the transverse momentum of the $j$th cluster, respectively. Furthermore, a Bayesian unfolding procedure~\cite{DAgostini:2010hil} in the RooUnfold software package~\cite{Adye:2011gm} is implemented to correct the jet mass, where the response matrix is obtained by embedding dijet-triggered p+p events into non-triggered Pb+Pb events. Our results are finally obtained from the unfolded jet mass distribution by unfolding the jet masses from Pb+Pb dijet events with the response matrix.

\section{Results and discussion} \label{3}

\subsection{Charged jet mass distribution}\label{3.1}

\begin{figure*}[htb]
\centering
\includegraphics
[width=16.5cm]{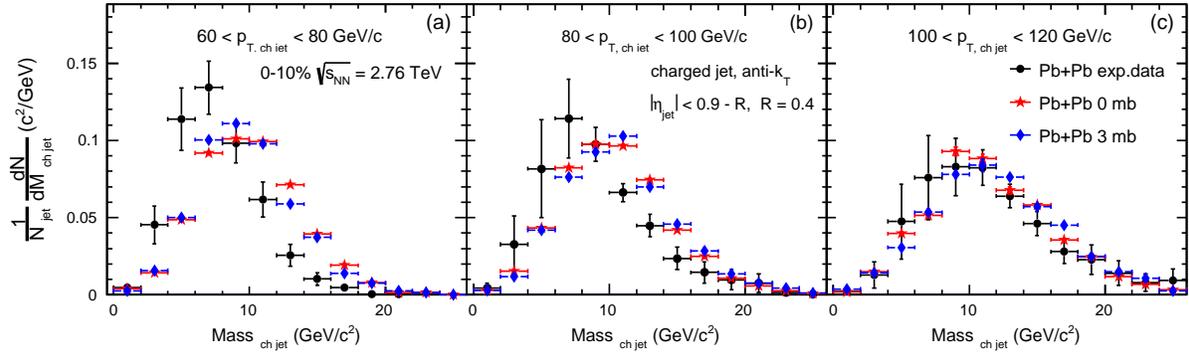}
\caption{The AMPT results on charged jet mass distributions for three $p_{\rm{T},\text{ch jet}}$ intervals in central Pb+Pb collisions (0-10\%) at 2.76 TeV, in comparison with the data from the ALICE experiment~\cite{ALICE_PLB776(2018)}.}
\label{fig.1}
\end{figure*}

Figure~\ref{fig.1} shows the AMPT results on the distributions of charged jet mass in three ranges of $p_{\rm{T},\text{ch jet}}$ in 0-10$\%$ most central Pb+Pb collisions at 2.76 TeV with a parton interaction cross-section of $0~\rm mb$ (only hadronic interactions) and $3~\rm mb$ (both partonic and hadronic interactions), in comparison with the data from the ALICE experiment~\cite{ALICE_PLB776(2018)}. We notice that the charged jet mass distributions (normalized per charged jet) from the AMPT simulations exhibit consistency without any apparent difference. The jet mass peaks are also consistent with that in p+p collisions [later shown in Figure~\ref{fig.4}(a)], which is due to the existence of Sudakov form factor in the jet mass distribution~\cite{Larkoski:2017fip}. Note that the suppression effect of Sudakov double logarithms becomes increasingly important, especially near the jet mass peak, which has driven many efforts to study the higher-logarithmic resummation of jet mass distribution~\cite{Dasgupta:2012hg,Chien:2012ur,Liu:2014oog,KZB_JHEP10(2018),Idilbi:2016hoa,Balsiger_JHEP04(2019)}. We observe that as the transverse momentum of charged jet increases, charged jet mass distribution becomes broader with a gradual increase of peak position. The results from the AMPT model overestimate the measured jet mass distributions for $60< p_{\rm{T},\text{ch jet}} < 80~\rm{GeV}$/$c$ and $80< p_{\rm{T},\text{ch jet}} < 100~\rm{GeV}$/$c$ in Figure~\ref{fig.1} (a) and (b), but well describe the experimental measurement for $100< p_{\rm{T},\text{ch jet}} < 120~\rm{GeV}$/$c$ in Figure~\ref{fig.1} (c). We find out that the discrepancy between the AMPT model and the experimental data for $60< p_{\rm{T},\text{ch jet}} < 100~\rm{GeV}$/$c$ in 0-10$\%$ central Pb $+$ Pb collisions actually comes from the corresponding discrepancy in p $+$ p collisions. Ones could adjust some PYTHIA parameters in HIJING part of the AMPT model to improve the description, and we would like to leave it as future work.

\begin{figure*}[htb]
\centering
\includegraphics
[width=16.5cm]{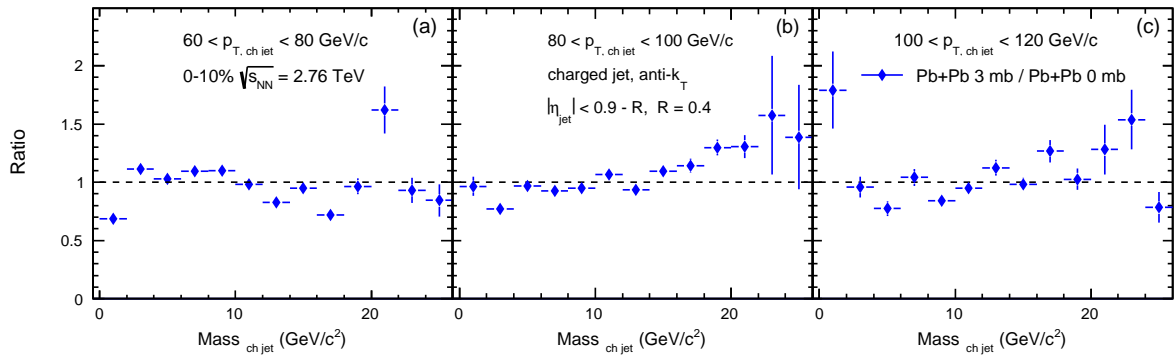}
\caption{The AMPT results on ratios of charged jet mass distribution of $3~\rm mb$ to $0~\rm mb$ for three $p_{\rm{T},\text{ch jet}}$ intervals in central Pb+Pb collisions (0-10\%) at 2.76 TeV.}
\label{fig.2}
\end{figure*}

To look for any effect of jet quenching on charged jet mass distributions, Figure~\ref{fig.2} (a)-(c) presents the ratios of charged jet mass distributions between with and without parton interactions in three ranges of $p_{\rm{T},\text{ch jet}}$ in central Pb+Pb collisions at 2.76 TeV from the AMPT model simulations. We found that if the data points with large errors on both edges are ignored, the ratios are consistent with around unity in the jet mass range of 5 - 20 $\rm{GeV^{2}}$/$c$, which means that no obvious effect from jet quenching effect is observed. In other words, the jet quenching effect due to parton interactions is so small that it is hard to be seen in the jet mass distributions in the final state of central Pb+Pb collisions from our simulations.

\begin{figure}[htbp]
\centering
\includegraphics
[width=7.5cm]{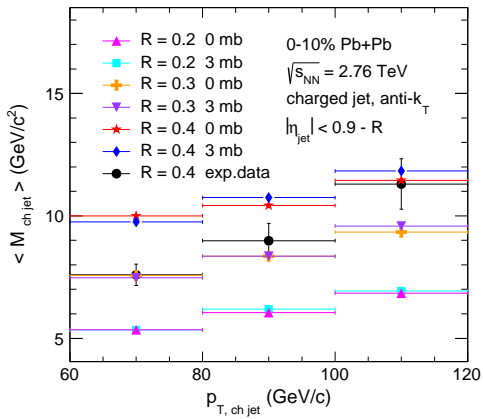}
\caption{The AMPT results on mean charged jet mass as a function of the transverse momentum of jet $p_{\rm{T},\text{ch jet}}$ and jet radius $R$ in central Pb+Pb collisions (0-10\%) at 2.76 TeV, in comparison with the ALICE data for $R$ = 0.4~\cite{ALICE_PLB776(2018)}.}
\label{fig.3}
\end{figure}

In order to further reveal the distinctions of charged jet mass between with and without partonic interactions, the mean charged jet mass is calculated as a function of $p_{\rm{T},\text{ch jet}}$ and $R$, as shown in Figure~\ref{fig.3}, where three cases of jet radius parameter $R$ = 0.2, 0.3, and 0.4 are compared. It is observed that the mean charged jet mass increases as the jet radius $R$ and the jet momentum $p_{\rm{T},\text{ch jet}}$ increase. Our result is in a similar trend to the experimental data for $R$ = 0.4. Our result only can describe the experimental data for the hardest $p_{\rm{T},\text{ch jet}}$ case, but overestimates the other two $p_{\rm{T},\text{ch jet}}$ cases. On the other hand, the difference between 3 mb and 0 mb looks most obvious for the jet with the largest jet radius $R$ with the hardest $p_{\rm{T},\text{ch jet}}$, which suggests that the hard jets with large jet cone sizes are more sensitive to the jet quenching effect than the soft jets with small jet cone sizes. These observations are also consistent with the expectation from Eq.(\ref{eq.2}).

\subsection{Stage evolution of full jet mass}\label{3.2}

\begin{figure*}[htb]
\centering
\includegraphics
 [width=16.5cm]{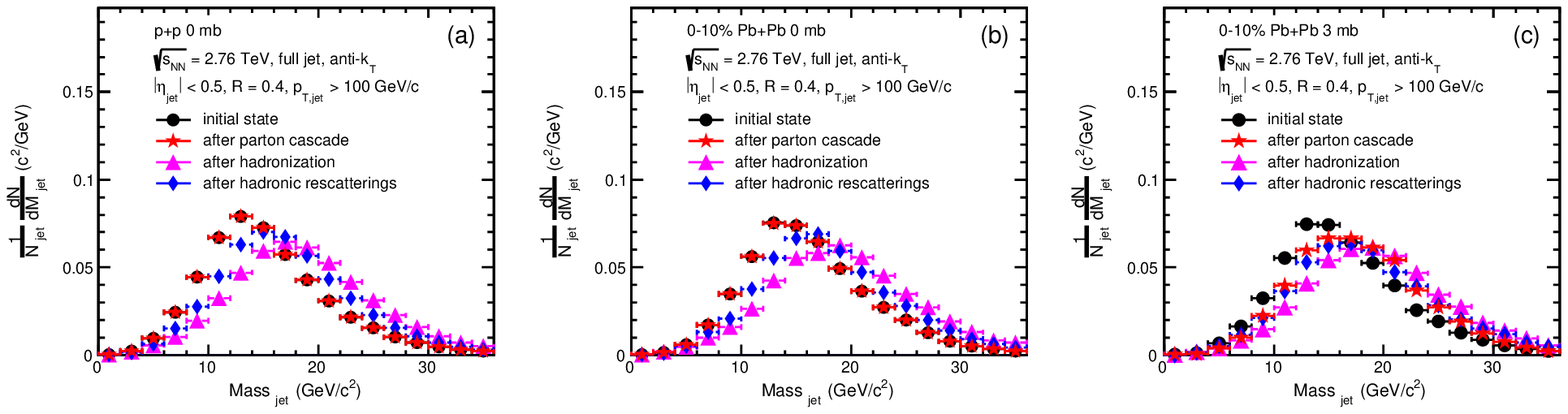}
\caption{The AMPT results on full jet mass distributions at different evolution stages in p+p collisions ($0~\rm mb$) at 2.76 TeV [plot (a)], and in central Pb+Pb collisions at 2.76 TeV [0 mb, plot (b) and 3 mb, plot (c)].
}\label{fig.4}
\end{figure*}

\begin{figure*}[htb]
\centering
\includegraphics
 [width=16.5cm]{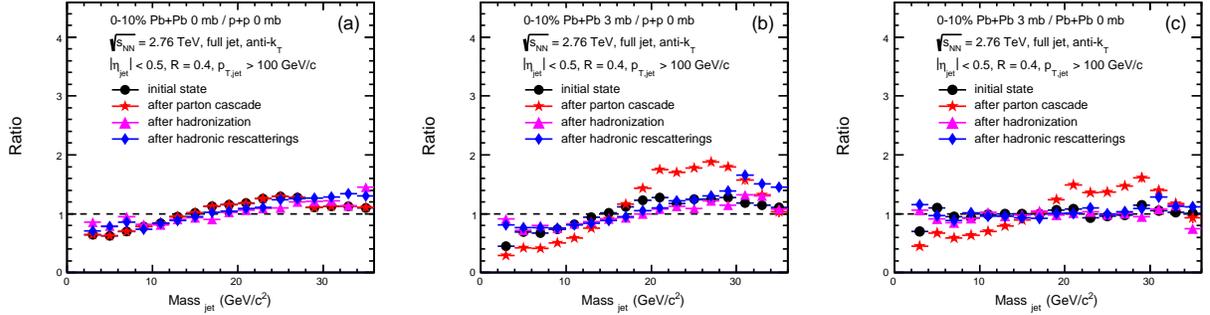}
\caption{
The AMPT results on the ratio of full jet mass distributions between Pb+Pb collisions ($3~\rm mb$) and  p+p collisions ($0~\rm mb$)  [plot (a)] ,  between Pb+Pb collisions ($3~\rm mb$) and  p+p collisions ($0~\rm mb$)  [plot (b)] , and between Pb+Pb collisions ($3~\rm mb$) and  Pb+Pb collisions ($0~\rm mb$) [plot (c)] for $p_{\rm{T},\text{jet}} > 100~\rm{GeV}$/$c$.
}\label{fig.5}
\end{figure*}

Relativistic heavy-ion collisions are actually a dynamical evolution that includes several important evolution stages. To characterize the medium effect on jet mass, it is essential to compare jet mass distributions from various stages of the AMPT model. We denote the four evolution stages as ``initial state'', ``after parton cascade'', ``after hadronization'', and ``after hadronic rescatterings'', respectively. We focus on the full jets which are reconstructed with both neutral and charged particles, in order to avoid the uncertainties of the conversion between neutral and charged particles among different evolution stages. Figure~\ref{fig.4} (a)-(c) present the distributions of full jet mass ($p_{\rm{T},\text{jet}} > 100~\rm{GeV}$/$c$) for four different evolution stages in p+p collisions and 0-10$\%$ most central Pb+Pb collisions with and without parton interactions from the AMPT simulations, respectively.
 
Because there should be almost no medium effect on jet mass in p+p collisions, the result from p+p collisions should be a good reference. For p+p collisions in Figure~\ref{fig.4} (a), we find that the full jet mass distribution (normalized per jet) in the initial state is basically as same as that that after parton cascade, since we turn off parton interactions with a zero cross section. Whereas, the hadronization process has an obvious influence on the jet mass. The jet mass distribution is shifted to a larger value via the hadronization. This kind of behavior that hadronization increases jet mass actually has been observed in the recent STAR experiment~\cite{STAR:2021lvw} also in PYTHIA-8~\cite{Sjostrand_CPCo178(2008)} simulations with hadronization turned on and off. It is consistent with the expectation from the non-perturbative effect that partons are confined into hadrons and obtain masses via hadronization. However, hadronic rescatterings result in a decrease of jet mass from the AMPT simulations, since both hadronic scatterings and resonances decays can change their kinematics to make some hadrons out of the jet cone.  The jet mass distributions in central Pb+Pb collisions with turning on and off parton interactions are presented in Figure~\ref{fig.4}(b) and Figure~\ref{fig.4}(c), respectively. In Figure~\ref{fig.4}(b), the full jet mass distribution in the initial state is basically as same as that after parton cascade, which is also similar to that observed in p+p collisions. Nevertheless, a right shift is observed from the initial state to after parton cascade if we turn on parton interactions in central Pb+Pb collisions, as shown in Figure~\ref{fig.4}(c). The shift of jet mass distribution indicates that the jet quenching effect in our model which includes both jet energy loss and medium response can lead to an increase of jet mass. It is consistent with our previous result of jet shape which shows more excited medium partons are involved into the jet cone due to parton cascade, especially contributing to the periphery of the jet cone~\cite{MGL_PRC89(2014)}. Unfortunately, the non-perturbative effect from hadronization and hadronic rescatterings weaken the right shift resulting from jet-medium interactions due to the same reason in p+p collisions.  This makes it difficult to observe significant hot medium modification effects on the plain jet mass distribution in the final state of relativistic heavy-ion collisions. Note that we apply the same $p_{\rm{T},\text{jet}}$ cut to jets from different evolution stages. However, we check that our conclusions will not change whether we choose the jets according to $p_{\rm{T},\text{jet}}$ in the initial or final stage.

To see the nuclear modification effects on jet mass distribution in heavy-ion collisions, we show the ratio of jet mass distribution in central Pb+Pb collisions with 0 and 3 mb to that in p+p collisions with 0 mb in Figure~\ref{fig.5} (a) and (b), respectively.  In Figure~\ref{fig.5} (a), we find that the ratio keeps different from the unity from the initial state to the final state. This indicates that the modification of jet mass distribution originates from the initial state, which may arise from the cold nuclear matter effect or different underlying event contribution in heavy-ion collisions. In Figure~\ref{fig.5} (b), we observe a strong modification of jet mass distribution after parton cascade, which is due to jet quenching in the hot QGP medium. However, the hot medium modification effect is weakened by the following process of hadronization, which ultimately leads to no significant effect of jet quenching in the final state. To see the difference of jet mass distributions between with and without parton interactions at different evolution stages, the full jet mass distribution ratios of 3 mb to 0 mb for four evolution stages in most central Pb+Pb collisions are shown in Figure.~\ref{fig.5}(c). It can be clearly seen that the ratio is different from unity after parton cascade, which indicates there is an increase of jet mass due to jet-medium interactions in parton cascade. However, the ratios become consistent with unity for the two final stages. It indicates that although the jet mass moves towards a higher value after jet passes through the QGP matter, unfortunately the discrepancy resulting from jet quenching is strongly reduced by hadronization. Therefore, the non-perturbative effects of hadronization and hadronic rescatterings make it difficult to observe the QGP medium modification on the plain jet mass. 

\begin{figure*}[htbp]
\centering
\includegraphics
[width=16.5cm]{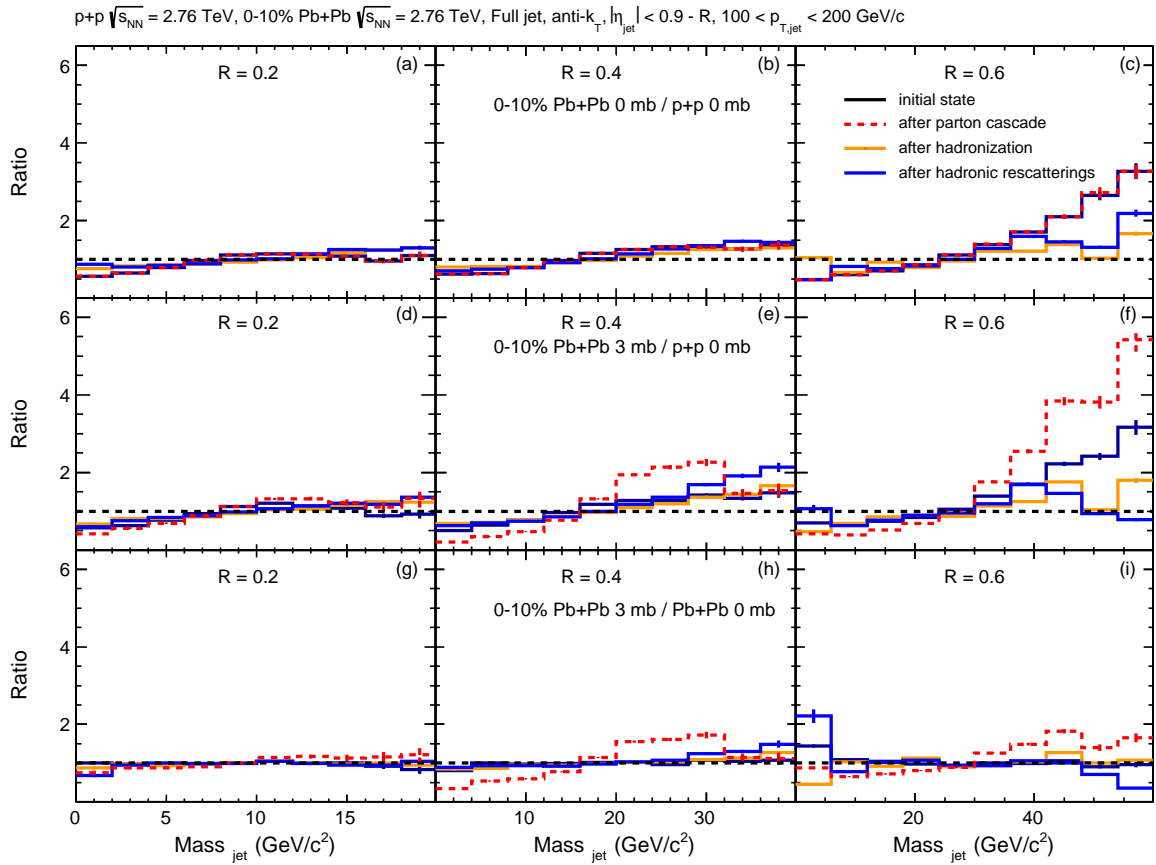}
\caption{Top panels: The AMPT results on the ratios of full jet mass distribution in 0-10$\%$ central Pb $+$ Pb collisions (0 mb) to that p $+$ p collisions at four different evolution stages, for $100< p_{\rm{T},\text{jet}} < 200~\rm{GeV}$/$c$ with three jet radii ($R$ = 0.2, 0.4, and 0.6)  [plots (a), (b), and (c)]. Middle panels: Same as the top panels but for the ratios of 0-10$\%$ central Pb $+$ Pb collisions (3 mb) to p $+$ p collisions [plots (d), (e), and (f)]. Bottom panels: Same as the top panels but for the ratios of 0-10$\%$ central Pb $+$ Pb collisions (3 mb) to Pb $+$ Pb collisions (0 mb) [plots (g), (h), and (i)].}
\label{fig.6}
\end{figure*}

To study the jet radius dependence of the modification of jet mass distribution, the ratios of full jet mass distributions for three jet radii (R=0.2, 0.4, and 0.6) at four evolution stages in 0-10$\%$ central Pb $+$ Pb collisions with 0 mb and 3 mb to that in p $+$ p collisions with 0 mb are shown in the top and middle panels in Figure~\ref{fig.6}. We observe the most obvious modification for the largest jet radius ($R$ = 0.6) at the initial state of central Pb $+$ Pb collisions, which indicates a larger cold nuclear matter effect or underlying event contribution for jets with larger jet radii. Meanwhile, we observe that the modification of jet mass distribution after parton cascade increases with increasing the jet radius in the middle panels in Figure~\ref{fig.6}. However, these modifications are the results of the combined effect of cold nuclear effect and jet quenching effect. To see the pure effect of jet quenching, we plot the full jet mass distribution ratios of 3 mb to 0 mb with three jet radii for four evolution stages in central Pb + Pb collisions in the bottom panels in Figure~\ref{fig.6}. There is the smallest effect of jet quenching after the parton cascade for the smallest jet radius ($R$ = 0.2). On the other hand, the jet quenching effect is found similar for the jet radii $R$ = 0.4 and 0.6. Unfortunately, the jet quenching effect is weakened by the process of hadronization, which eventually leads to no significant effect of jet quenching after hadronic rescatterings. 

Note that although only the jet elastic energy loss is considered in our work, we expect that the qualitative effect of the dynamic evolution on the jet mass distribution will not change. In addition, we recently found that a new hybrid hadronization mechanism with both coalescence and fragmentation processes can better describe the transverse momentum $j_T$-dependent jet fragmentation functions in p+p and p+Pb collisions than the simple quark coalescence mechanism in the current AMPT model~\cite{Duan:2022bew}. Thus, it is interesting to study hadronization effects on the plain jet mass in both small and large colliding systems in the future. On the other hand, we expect that the groomed jet mass could provide a more powerful probe to explore the jet quenching effect ~\cite{CMS:2018fof,CSS_RPP84(2021)} in our future study, because the groomed jet mass has less contamination from non-perturbative effects~\cite{Dasgupta:2013ihk,Hoang:2019ceu}. 

On the other hand, we examine the distributions of quark-initiated and gluon-initiated jet mass distributions separately. We observe that the gluon-initiated jets generally have larger jet masses than quark-initiated jets due to the larger color charge and Casimir color factor $C_{\rm A} > C_{\rm F}$~\cite{Chien:2018dfn,Ying:2022jvy}. However, the pattern of changes in the mass distribution of quark-initiated and gluon-initiated jets is similar after each evolution stage. Finally, the average jet mass induced by gluons is larger than the average jet mass induced by quarks. Therefore, the average value and shape of the inclusive jet mass distribution should also be sensitive to the ratio of quark-initiated jets to gluon-initiated jets. This has actually attracted a great deal of experimental effort to distinguish quark-initiated jets from gluon-initiated jets~\cite{Metodiev:2018ftz,Komiske:2022vxg,ALICE-USA:2022glt}. However, this is beyond the scope of this paper, and we would like to leave it for our future study.

\section{Summary}\label{4}

In summary, the plain jet mass distributions of reconstructed jets have been investigated in p+p and 0-10$\%$ most central Pb+Pb collisions at $\sqrt{s_{\rm NN}} = 2.76~{\rm TeV}$ using a multiphase transport model with a string melting mechanism. Our results of charged jet mass distribution using jet radius $R$ = 0.4 can describe the experimental data for 100 $< p_{\rm{T},\text{ch jet}} < 120~\rm{GeV}$/$c$, but overestimate the charged jet mass for smaller ranges of $p_{\rm{T},\text{ch jet}}$. The mean charged jet mass depends strongly on jet transverse momentum and jet radius. The mean charged jet mass is observed as an increasing function of $p_{\rm{T},\text{ch jet}}$ and $R$ in central Pb+Pb collisions. It indicates that the hard jets with large jet cone sizes are more sensitive to the jet quenching effect than the soft jets with small jet cone sizes. 

Meanwhile, the full plain jet mass distributions for various dynamical evolution stages in relativistic heavy-ion collisions have been studied in detail. In central Pb+Pb collisions, the jet mass distribution is shifted to a higher value by parton cascade due to the jet quenching effect. Nevertheless, the discrepancy between with and without jet quenching effect is strongly diminished by non-perturbative effects from hadronization and hadronic rescatterings. The interplay of jet quenching and non-perturbative effects makes it difficult to observe significant effects of jet quenching on the plain jet mass distribution in the final state of relativistic heavy-ion collisions. We expect that the groomed jet mass could provide a more powerful probe to explore the jet quenching effect in our future study, because the groomed jet mass has less contamination from non-perturbative effects. 

\begin{acknowledgments}
We thank Prof. D.Y. Shao for the helpful discussions. This work is supported by the National Natural Science Foundation of China under Grants No.12147101, No. 11890714, No. 11835002, No. 11961131011, No. 11421505,  the National Key Research and Development Program of China under Grant No. 2022YFA1604900, the Strategic Priority Research Program of Chinese Academy of Sciences under Grant No. XDB34030000, and the Guangdong Major Project of Basic and Applied Basic Research under Grant No. 2020B0301030008. 

\end{acknowledgments}


\bibliography{reference}


\end{document}